\documentclass[12pt]{article}
\usepackage{amsmath}
\usepackage{amssymb}
\usepackage{graphicx}
\newcommand{\be}{\begin{equation}}
\newcommand{\ee}{\end{equation}}
\newcommand{\ba}{\begin{eqnarray}}
\newcommand{\ea}{\end{eqnarray}}

\newcommand{\lra}{{\small $\leftrightarrow$}}

\begin{document}
\vspace{1cm}

\begin{center}
{\bf\large Amplification of the induced ferromagnetism\\
in diluted magnetic semiconductor}
\end{center}
  \vspace{1cm}

\centerline{E.Z. Meilikhov\footnote{$^)$ e-mail:
meilikhov@imp.kiae.ru}$^)$, R.M. Farzetdinova}
\medskip
\centerline{\small\it Kurchatov Institute, 123182 Moscow, Russia}
  \vspace{1cm}

  \centerline{
\begin{tabular}{p{15cm}}
\footnotesize\qquad Magnetic properties of the planar structure
consisting of a ferromagnetic metal and the diluted magnetic
semiconductor are considered (by the example of the structure
Fe/Ga(Mn)As, experimentally studied in~\cite{1}). In the framework
of the mean field theory, we demonstrate the presence of the
significant amplification of the ferromagnetism, induced by the
ferromagnetic metal in the near-interface  semiconductor area, due
to the indirect interaction of magnetic impurities. This results in
the substantial expansion of the temperature range where the
magnetization in the boundary semiconductor region exists, that
might be important for possible practical applications.
\end{tabular}
} \vspace{0.5cm}

\centerline{\bf Introduction} \medskip

One of the major roadblocks on the way to developing the
semiconductor spintronics is now the lack of semiconductor materials
and structures that would be ferromagnetic at the room temperature.
Today, the record value of the Curie temperature is reached with the
diluted magnetic semiconductor Ga$_{1-x}$Mn$_x$As ($x=0.05-0.07$)
and equals $T_{\rm C}\approx 180$ K~\cite{2}. In this connection,
the interest has been aroused in hybrid structures of the
metal/semiconductor type whose magnetic properties are to the great
extent defined by the high-temperature ferromagnet\-ism of the metal
being the part of the structure. For instance, in the recent
paper~\cite{1} (selected for a Viewpoint in {\it Physics}) the
planar structure Fe/Ga$_{1-x}$Mn$_x$As(100) has been investigated.
It has been shown that the induced ferromagnetic state in the
near-interface layer of the semiconductor film survives even at the
room temperature. The magnetization of that semiconductor region is
opposite (by the direction) to the magnetization of Fe-film, that
suggests the antiferromagnetic exchange interaction between  Fe
atoms and impurity Mn atoms in GaAs matrix. The intensity of that
interaction is determined by the relevant effective magnetic field
$H_{\rm Fe}$ depending on the mutual disposition of a given Mn atom
and different Fe atoms interacting with the latter. In addition,
there is the indirect exchange interaction of the ferromagnetic
nature between Mn atoms putting into effect via mobile charge
carriers (for example, by the RKKY mechanism). That interaction is
characterized by the effective exchange field $H_{\rm Mn}$ depending
on the distance ${r_k}$ ($k=1,2,\ldots$) between a given Mn atom and
different Mn atoms interacting with the latter. Therewith, it is
essential that the carrier mobility in the diluted magnetic
semiconductor is so low (on the order of 10
cm$^2$/V$\cdot$c~\cite{3,4}) that due to the collision broadening of
the hole energy levels the system becomes to be effectively
three-dimensional and there is no need to take into account any
effect of the size quantization~\cite{5}.

We show that the interference of both mentioned interactions,
responsible for the magnetic ordering of Mn atoms, multiplies the
effect of each mechanisms severally. In fact, the magnetic "seeding"
which is the spin polarization of semiconductor magnetic atoms,
induced by the ferromagnet, is significantly amplified due to the
indirect interaction between those atoms. As a result, the Curie
temperature, determining the existence of the ferromagnetism in the
boundary region of the magnetic semiconductor increases essentially.
\vspace{0.5cm}

\centerline{\bf Mean field approach} \medskip

The magnetization in the boundary layer of the diluted magnetic
semiconductor is parallel to the interface plane and significantly
non-uniform along the direction perpendicular to the latter
($z$-axis). It is convenient to characterize such a non-uniform
magnetization by the reduced local magnetization $-1\leqslant
j({h})\equiv M({h})/M_s\eqslantless1$ ($h$ is the distance from the
interface plane, $M_s$ is the saturation magnetization). In the
framework of the mean field theory, it is defined by the equation
 \be\label{1}
j(h)=\tanh\left[\frac{\mu H_{\rm eff}(h)}{kT}\right],
  \ee
where $\mu$ is the magnetic moment of the impurity Mn atom,
 \be\label{2}
H_{\rm eff}(z)=\sum\limits_i H_{\rm Fe}({\bf R}_{i})+\sum\limits_k
H_{\rm Mn}(r_{k})
  \ee
is the total effective field effecting on a given Mn atom. Summation
is performed (in the first sum) over all Fe atoms, and (in the
second sum) over all Mn atoms.

When calculating the first sum in (\ref{2}) one should take into
account that considered Fe and Mn atoms are, in general,  situated
in media with different  exchange interaction lengths. Hence, the
result of that interaction is not a simple function of the distance
$R_i$ between those atoms. However, lest the calculation be
complicated by unprincipled details, we proceed below from the
simple exponential spatial dependence of the form
 \be\label{3}
\mu H_{\rm Fe}(R_{i})=j_{\rm Fe}({\bf R}_i)J_{\rm
Fe}\exp[-R_i/\ell_{\rm Fe}],
  \ee
where $J_{\rm Fe}$ and $\ell_{\rm Fe}$ are, accordingly, the
characteristic energy and the length of the interaction, $j_{\rm
Fe}({\bf R}_i)$ is the local magnetization of the Fe film in the
point of the relevant Fe atom location. Choosing the coordinate
origin in the interface plane, assume Fe layer being disposed in the
region $-L_{\rm Fe}<z<0$, and the semiconductor film is in the
interval $0<z<L_{\rm Mn}$. Then, in the continual approximation
 \be\label{4}
\sum\limits_i \mu H_{\rm Fe}(R_{i})= J_{\rm Fe}\,n_{\rm
Fe}\!\!\!\int\limits_{z=-L_{\rm Fe}}^0\int\limits_{\rho=0}^\infty j_{\rm
Fe}(z)\exp\left[-\sqrt{\rho^2+(h-z)^2}/\ell_{\rm Fe}\right]\!2\pi\rho d\rho dz,
  \ee
where $n_{\rm Fe}$ is the concentration of Fe atoms. In principle,
the magnetization $j_{\rm Fe}$ depends on the temperature, however
if the latter is significantly lower than the ferromagnetic Curie
temperature and the film itself is in the single-domain state, then
$j_{\rm Fe}(z)\equiv 1$, and it follows from (\ref{4})
 \be\label{5}
\sum\limits_i \mu H_{\rm Fe}(R_{i})= 4\pi n_{\rm Fe}\ell_{\rm
Fe}^3J_{\rm Fe}\Phi(h),\quad \Phi(h)=e^{-h/\ell_{\rm
Fe}}\left[1+\frac{h}{2\ell_{\rm Fe}}-\left(
 1+\frac{L_{\rm Fe}+h}{2\ell_{\rm Fe}}\right)e^{-L_{\rm Fe}/\ell_{\rm Fe}} \right].
  \ee

Before calculating the second sum in (\ref{2}) it should be noted
there is no complete understanding the nature of the ferromagnetism
in diluted magnetic semiconductors so far. Among  mechanisms leading
to the ferromagnetic ordering of magnetic impurity spins various
forms of their indirect interaction, induced by mobile charge
carriers, are considered: RKKY-exchange~\cite{6}, kinematic
exchange~\cite{7,8}, etc.~\cite{7}. In addition, there is the
universal Bloembergen-Rowland mechanism~\cite{9}, which does not
require the presence of mobile charge carriers (or their high
concentration leading to the carrier degeneration) and could be drew
for describing ferromagnetism in systems of the  Ga(Mn)As, Ga(Mn)N
type~\cite{10}. The essence of that mechanism is the interaction of
impurity spins via virtual holes appearing in the valence band on
the transition of electrons from that band to acceptor levels. The
energetic threshold of this process, concerned with the finite
energy $\Delta$ of the acceptor ionization, results in the
exponential decay of the relevant interaction with the
inter-impurity distance. The characteristic length of the decay is
defined by the de Broglie  wave length
$\lambda_\Delta=\hbar(2m\Delta)^{-1/2}$ of holes with the energy
$\Delta$ and the mass $m$. The expression for the effective exchange
field $H_{\rm eff}$ of the indirect interaction of two parallel
impurity spins separated by the distance $r$ reads~\cite{10}
 \be\label{a}
\mu H_{\rm eff}(r)=-\frac{J_{pd}^2m^2\Delta}{\pi^3\hbar^4 N^2
r^2}K_2(2r/\lambda_\Delta),
 \ee
where $J_{pd}$ is the energy of the contact interaction of the
impurity spin with a hole, $N$ is the concentration of matrix atoms,
$K_2$ is the McDonald function. For long distances ($r\gtrsim
\lambda_\Delta/2$),  $K_2(2r/\lambda_\Delta)\approx
(4r/\pi\lambda_\Delta)^{-1/2}\exp(-2r/\lambda_\Delta)$
asymptotically. Keeping for that case in (\ref{a}) the most
essential exponential part of the interaction spatial dependence
only, one writes it in the form
 \be\label{b}
\mu H_{\rm eff}(r)=-J_0\exp(-r/\ell),
 \ee
ãäå $\ell=\lambda_\Delta/2$, $J_0\sim J_{pd}^2/8\pi^3\Delta
N\lambda_\Delta^3$.

Similar expression for the energy of the magnetic impurity
interaction is exploited when investigating the kinematic exchange
model~\cite{7}. At last, if one turns to the RKKY interaction, under
the strong scattering of carriers (see above) its intensity drops
with the distance so rapidly that one could neglect the
alternating-sign nature of that interaction and describe it by means
the main exponential factor of the form $\exp(-r/\ell)$ ($\ell$ is
the length of the spin coherency).

In the light of the above-stated we use the following model relation
for the indirect interaction of a given Mn atom with another Mn atom
situated in the point with the local magnetization equal to
$j(z_k)$:
 \be\label{6}
\mu H_{\rm Mn}(r_{k}, z_k)=j(z_k)J_{\rm Mn}\exp[-r_k/\ell_{\rm Mn}].
  \ee
Here $J_{\rm Mn}$ and $\ell_{\rm Mn}$ are, accordingly, the
characteristic energy and the length of the indirect interaction. In
the continual approximation,
 \ba\label{7}
\sum\limits_k \mu H_{\rm Mn}(r_{k}, z_k)= J_{\rm Mn}\,n_{\rm
Mn}\!\!\!\int\limits_{z=0}^{L_{\rm Mn}}\int\limits_{\rho=0}^\infty
j(z)\exp\left[-\sqrt{\rho^2+(h-z)^2}/\ell_{\rm Mn}\right]\!2\pi\rho d\rho dz=\nonumber\\
=2\pi n_{\rm Mn}\ell_{\rm Mn}^3J_{\rm Mn}\int\limits_0^{L_{\rm Mn}}
K(z,h)j(z)dz,\hspace{110pt}
  \ea
where  $n_{\rm Mn}$ is the concentration of Mn atoms,
 \be\label{8}
K(z,h)=\left(1+\frac{|h-z|}{\ell_{\rm
Mn}}\right)\frac{e^{-|h-z|/\ell_{\rm Mn}}}{\ell_{\rm
Mn}}.\hspace{55pt}
 \ee
Unlike the infinite system, the value of the integral (\ref{7})
depends on the coordinate $h$ of chosen point.

The mean field equation (\ref{1}) could be now written in the form
 \be\label{9}
j(h)=\tanh\left[\frac{1}{\tau}\left(c_{\rm Fe}\Phi(h)+c_{\rm
Mn}\int\limits_0^{L_{\rm Mn}} K(z,h)j(z)dz \right)\right],
 \ee
where $c_{\rm Fe}=4\pi n_{\rm Fe}\ell_{\rm Fe}^3 (J_{\rm Fe}/J_{\rm
Mn})$, $c_{\rm Mn}=2\pi n_{\rm Mn}\ell_{\rm Mn}^3 $ are structure
parameters, $\tau=kT/J_{\rm Mn}$ is the reduced temperature. That is
the non-linear integral equation determining the profile and the
temperature dependence of the magnetization in the semiconductor
part of the structure.  It is non-local: the magnetization $j(h)$ is
defined by all points ($0<z<L$) of the Ga(Mn)As film.\vspace{0.5cm}

\centerline{\bf Results} \medskip

Numerical solution of Eq. (\ref{9}) has been found, in the same way
as in~\cite{5}, by the successive-approximation method. Though we
have kept in mind the concrete structure
Fe/Ga$_{1-x}$Mn$_x$As~\cite{1}, the qualitative character of our
examination makes using exact values of parameters, governing the
system behavior, to be surplus.  Therefore, we accept that the
constant $a$ of the (face centered cubic) Fe lattice is about half
as the constant of the (body centered cubic) sublattice of Ga atoms
(which are replaced by Mn atoms) and set $n_{\rm Fe}a^3=2$, $n_{\rm
Mn}a^3=0.05$ (that corresponds to $x\approx0.1$), and for other
parameters accept the values $L_{\rm Fe}=L_{\rm Mn}=7a$~\cite{1},
$\ell_{\rm Fe}=a$, $\ell_{\rm Mn}=1.5a$. As for the ratio $J_{\rm
Fe}/J_{\rm Mn}$, it has been varied over a wide range (see below).

Putting $c_{\rm Fe}=0$ in (\ref{9}),  one could find the temperature
interval of the "intrinsic" (non-induced by the Fe film)
ferromagnetism in Ga(Mn)As. The relevant Curie temperature occurs to
be equal $\tau_{\rm C}\approx 5$, or $kT_{\rm C}=5J_{\rm Mn}$ (see
below Fig. 2). For that case, in Fig. 1 (left panel, curve Mn\lra
Mn) the spatial distribution $j_{\rm Mn}(z)$ of the local Mn
magnetization at the temperature  $\tau=4<\tau_{\rm C}$ is shown. As
one would expect, it is symmetric about the middle plane ($z=L/2$)
of the semiconductor layer. In the same Fig. 1, one could see the
spatial magnetization distribution of Mn atoms, induced by the their
exchange interaction with Fe atoms only, without intrinsic indirect
interaction (left panel, curve Mn\lra Fe) for $J_{\rm Fe}/J_{\rm
Mn}=2.5$. Such an induced magnetization drops rapidly with moving
away from the interface boundary Fe/Ga(Mn)As ($z=0$). At last, the
third curve (Mn\lra Mn+Mn\lra Fe) in the left panel of Fig. 1 is the
result of the combined action of both mechanisms of  magnetic
ordering  manifesting in the high amplification of the induced
ferromagnetism by the indirect interaction  of magnetic Mn
impurities.  The respective magnetization gain is especially high
away from the boundary Fe/Ga(Mn)As ($z\gtrsim L/2$), where the
induced magnetization is enlarged up to fivefold value.

It is remarkable that the substantial amplification of the induced
magnetization remains even at temperatures being much higher the
Curie temperature corresponding to the intrinsic ferromagnetism of
Ga(Mn)As. The middle panel in Fig. 1, referring to the temperature
$\tau=6>\tau_{\rm C}$, demonstrates that the gain of the induced
ferromagnetism even in this case remains equally high, though away
from the interface the absolute value of the magnetization drops
somehow.

The magnetization gain in magnetic semiconductor stimulated by the
spatially localized  "magnetic seeding" is obviously demonstrated in
the right panel of Fig. 1, where as the spatial dependence of the
effective exchange interaction of Fe and Mn atoms we have chosen the
linear approximation $\Phi(0)+\Phi'(0)h=1-h/2\ell_{\rm Fe}$ of the
function (\ref{5}) (different from zero in the range $0<h<2\ell_{\rm
Fe}$ only). It is seen that even at the temperature $\tau=6$, which
is essentially exceeds the Curie temperature $\tau_{\rm C}\approx4$,
magnetic ordering appears even in that region of the magnetic
semiconductor where the induced seed magnetization is absent.

It is convenient to characterize the non-uniformly magnetized layer
Ga(Mn)As by the average magnetization
 \be\label{10}
\langle j_{\rm Mn}\rangle=\frac{1}{L}\int\limits_{0}^{L}j_{\rm Mn}(z)dz.
 \ee
Temperature dependencies of that value for the intrinsic (Mn\lra
Mn), induced (Mn\lra Fe) and amplified combined (Mn\lra Mn+Mn\lra
Fe) ferromagnetism are shown in Fig. 2 (left panel: $\ell_{\rm
Mn}=1.5a$, $J_{\rm Fe}/J_{\rm Mn}=0.5$, middle panel: $\ell_{\rm
Mn}=1.5a$; $J_{\rm Fe}/J_{\rm Mn}=5$). It is seen that significant
magnetization of Ga(Mn)As remains even at the temperature which is
by six (and more) times higher than the intrinsic Curie temperature.
Such a giant expansion of the temperature interval of existing the
ferromagnetism near the interface Fe/Ga(Mn)As has been
experimentally obser\-ved in~\cite{1}.

The effect of amplifying the induced ferromagnetism in the
considered system due to the indirect interaction of magnetic
impurities in the semiconductor depends  severely on the intensity
$J_{\rm Fe}$ of the exchange interaction Mn\lra Fe and temperature.
Corresponding dependencies are shown in Fig. 3, wherefrom it is seen
that though the maximum relative effect (defined by the ratio
$(j_{\rm MM+MF}-j_{MF})/j_{\rm MF}$, where $j_{MF}$, $j_{\rm MM+MF}$
are magnetisations corresponding to Mn\lra Fe and Mn\lra Fe+Mn\lra
Mn curves) is observed at a temperature somewhat lower than the
intrinsic Curie temperature (in the considered case, at
$\tau\approx\tau_{\rm C}\approx5$), the significant (and the most
important) amplification effect  remains even at much higher
temperatures.

At last, notice that varying other system parameters does not lead
to the qualitative modification of results, describing the
properties of the considered system. For instance, three temperature
dependencies $\langle j_{\rm Mn}(\tau)\rangle$ of the
above-mentioned average magnetisations for the case $\ell_{\rm
Mn}=3a$, $J_{\rm Fe}/J_{\rm Mn}=0.5$, shown in Fig.~ 1 (right
panel), differ from the same dependencies for the case $\ell_{\rm
Mn}=1.5a$, $J_{\rm Fe}/J_{\rm Mn}=0.5$ (see the left panel) mainly
by the scale change. Conducted calculations demonstrate the
"stability" of basic results relative to varying problem
parameters.\bigskip

\centerline{\bf Conclusions} \medskip

Magnetic properties of the planar structure consisting of a
ferromagnetic metal and the diluted magnetic semiconductor are
considered (by the example of the structure Fe/Ga(Mn)As,
experimentally studied in~\cite{1}). In the framework of the mean
field theory, we demonstrate the presence of the significant
amplification of the ferromagnetism, induced by the ferromagnetic
metal in the near-interface  semiconductor area, due to the indirect
interaction of magnetic impurities. This results in the substantial
expansion of the temperature range where the magnetization in the
boundary semiconductor region exists -- far beyond the limits of the
interval limited from above by the Curie temperature of the magnetic
semiconductor itself. Results might be used for describing
properties of combined nanosized systems ferromagnetic/diluted
magnetic semiconductor. \bigskip

This work has been supported by Grant \# 09-02-00579 of the Russian
Foundation of Basic Researches.

\newpage
\renewcommand{\refname}{\centerline{\normalsize\bf References}\vspace{5mm}}

\newpage
\centerline{\bf Figure captions}
\bigskip
\bigskip

Fig. 1. Spatial distributions of the local semiconductor
magnetization $j_{\rm Mn}(z)$  near the interface Fe/Ga(Mn)As at
temperatures $\tau=4<\tau_{\rm C}$ (left panel) and
$\tau=6>\tau_{\rm C}$ (middle and right panels) for $J_{\rm
Fe}/J_{\rm Mn}=2.5$. Mn\lra Mn is the intrinsic magnetization of the
semiconductor, Mn\lra Fe is induced one, Mn\lra Mn+Mn\lra Fe is the
induced magnetization amplified by the indirect interaction. The
spatial dependence of the effective exchange interaction of Fe and
Mn atoms: left and middle panels -- Eq.(\ref{5}), right panel -- the
linear approximation (see text).\\

Fig. 2. Temperature dependencies of the average magnetization
$\langle j_{\rm Mn}\rangle$ for the intrinsic (Mn\lra Mn), induced
(Mn\lra Fe) and amplified combined (Mn\lra Mn+Mn\lra Fe)
ferromagnetism of the considered structure. Left panel: $\ell_{\rm
Mn}=1.5a$, $J_{\rm Fe}/J_{\rm Mn}=0.5$, middle panel: $\ell_{\rm
Mn}=1.5a$, $J_{\rm Fe}/J_{\rm Mn}=5$, right panel: $\ell_{\rm
Mn}=3a$, $J_{\rm Fe}/J_{\rm Mn}=0.5$. \\

Fig. 3. Dependencies of the average magnetization $\langle j_{\rm
Mn}\rangle$ for the induced (Mn\lra Fe) and amplified combined
(Mn\lra Mn+Mn\lra Fe) ferromagnetism of the considered structure on
the exchange (Mn\lra F) interaction  energy $J_{\rm Fe}$ at the
temperature $\tau=6>\tau_{\rm C}$. In the insert -- temperature
dependencies of the relative gain effect for the induced
magnetization at various  $J_{\rm Fe}$ values.\\


\begin{thebibliography}{30}


\bibitem{1} F. Maccherozzi, M. Sperl, G. Panaccione, J. Mina'r,
S. Polesya, H. Ebert, U. Wurstbauer, M. Hochstrasser, G. Rossi, G.
Woltersdorf, W. Wegscheider, C. H. Back, Phys. Rev. Lett., {\bf
101}, 267201 (2008).

\bibitem{2}  M. Wang, R. P. Campion, A. W. Rushforth, K.
W. Edmonds, C. T. Foxon, B. L. Gallagher, Appl. Phys. Lett. {\bf
93}, 132103 (2008).

\bibitem{3} D. Neumaier, M. Schlapps, U. Wurstbauer, J. Sadowski, M.
Reinwald, W. Wegscheider, D. Weiss, arXiv:0711.3278v2 (2007).

\bibitem{4} R. Moriya, H. Munekata, J. Appl. Phys. {\bf 93}, 4603
(2003).

\bibitem{5}  E.Z. Meilikhov, R.M. Farzetdinova, IEEE Trans. on Magn., {\bf 44},
2871 (2008).

\bibitem{6} T. Dietl, H. Ohno, F. Matsukara, J. Cibert, and D. Ferrand, Science, \textbf{287},
             1019 (2000).

\bibitem{7}T. Jungwirth, Jairo Sinova, J. Ma\v{s}ek, J. Kuc\v{e}ra, A. H.
MacDonald, Rev. Mod. Phys. \textbf{78}, 809 (2006); exhaustive
bibliography on this subject is  placed on the site
http:{//}unix12.fzu.cz/ms/navigate.php?cont=public\_\,in.

\bibitem{8} P.M. Krstajic, V.A. Ivanov, F.M. Peeters, V.Fleurov, and K. Kikoin,
Europhys. Lett., \textbf{61}, 235 (2003).

\bibitem{9} N. Bloembergen, and T.J. Rowland, Phys. Rev., {\bf 97}, 1679 (1955).

\bibitem{10} V.I. Litvinov, and V.K. Dugaev, Phys. Rev. Lett., {\bf 86}, 5593 (2001).

\end{thebibliography}
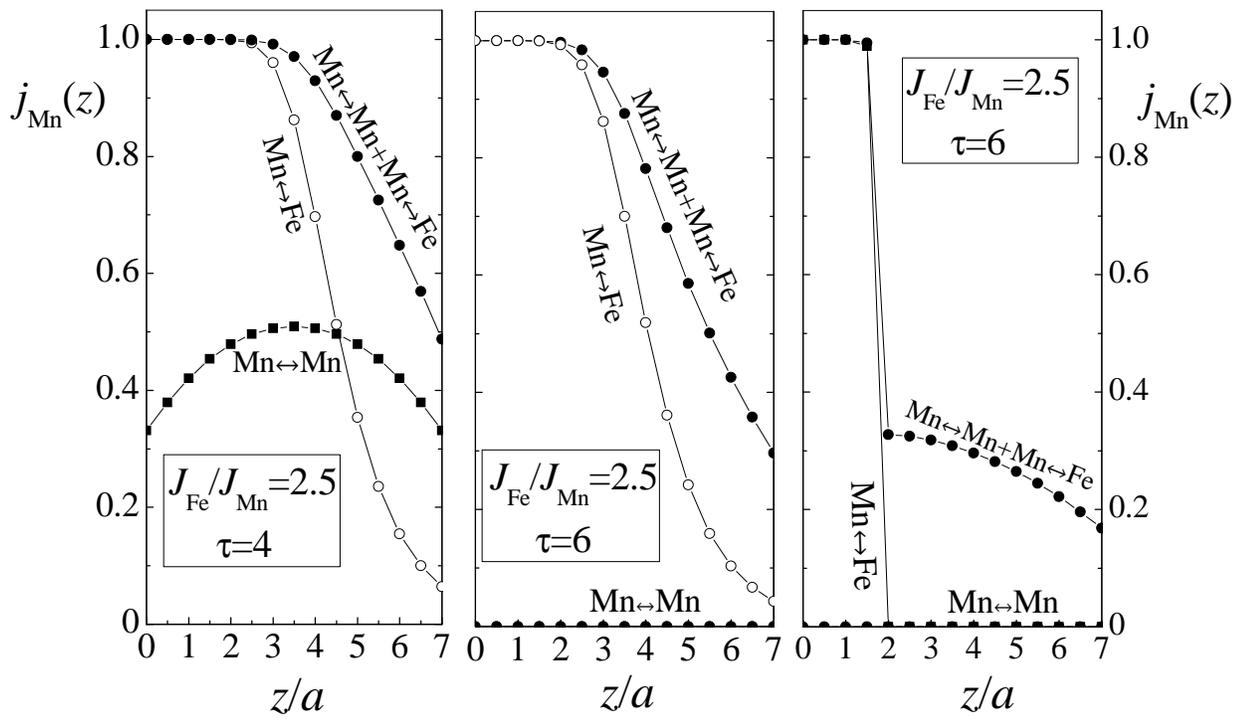
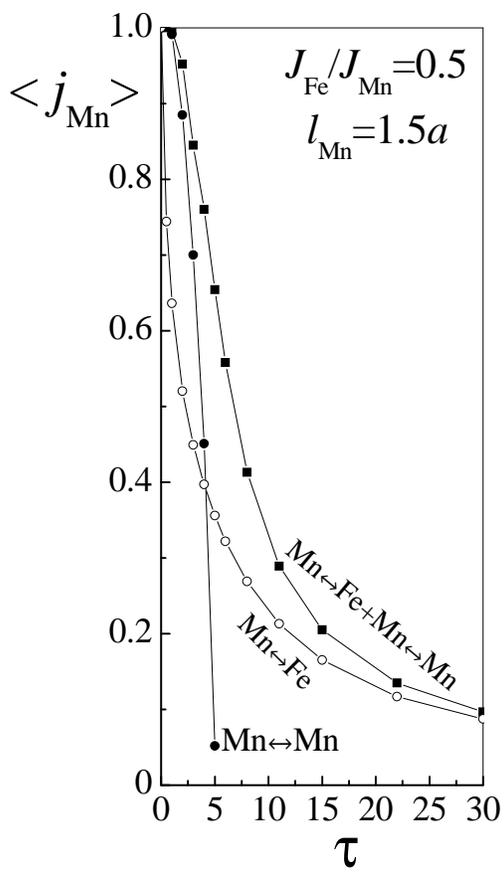
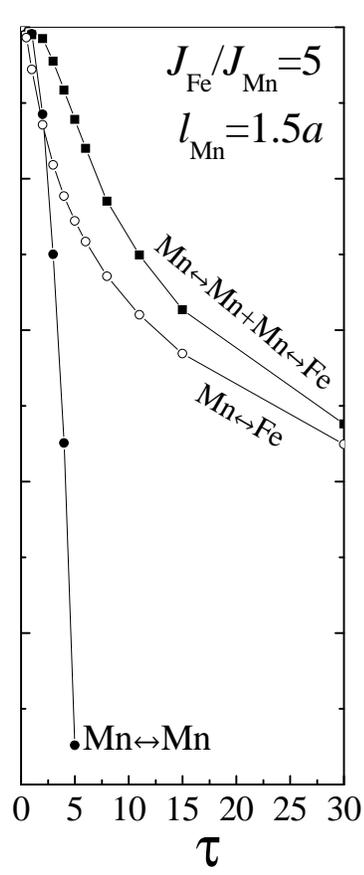
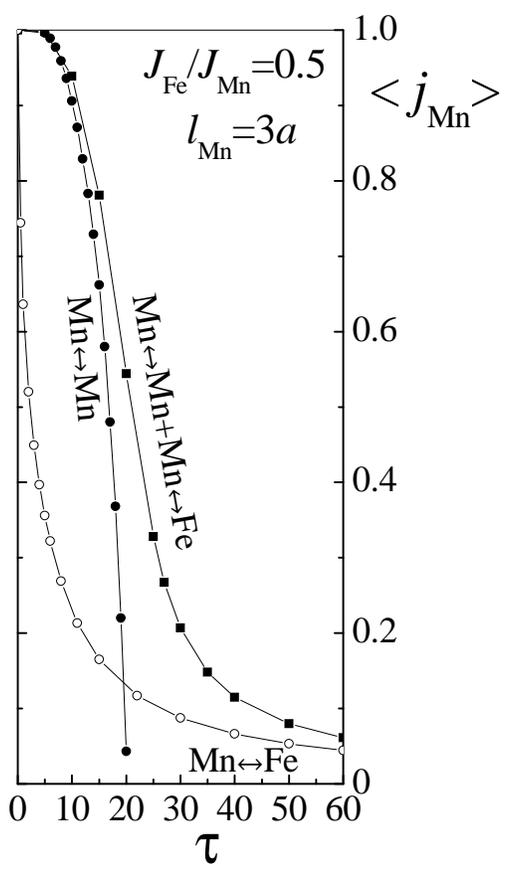
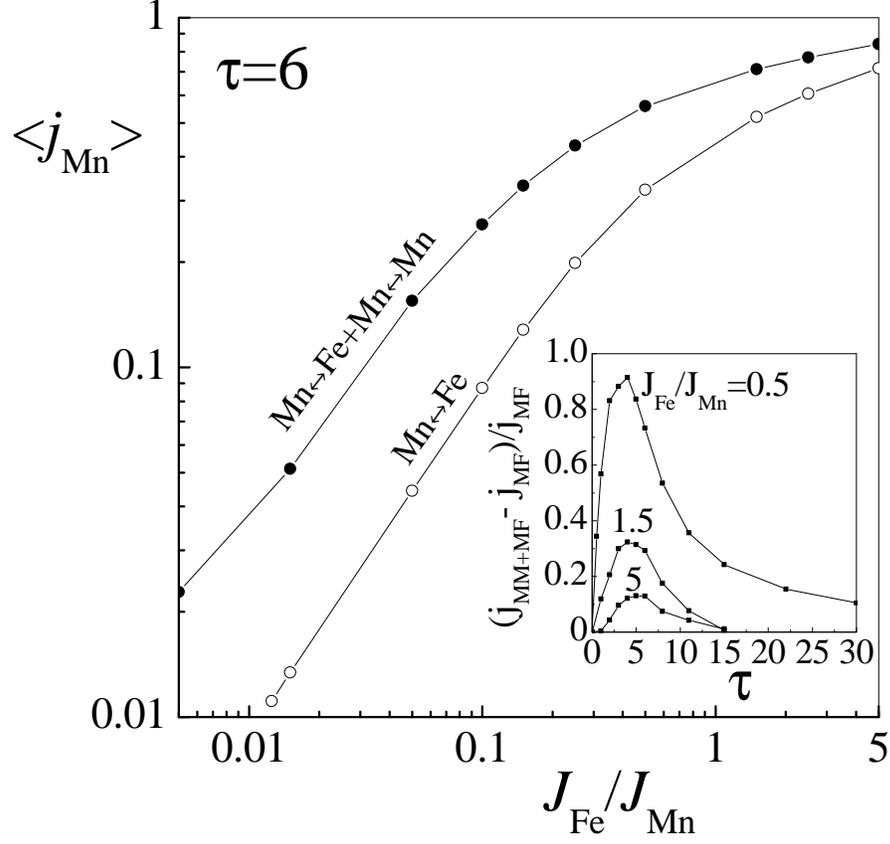
\end{document}